\documentclass[twoside,epsfig,psfig]{article}

\usepackage{fleqn,espcrc2}
\def\lsim{\mathrel{\rlap{\lower4pt\hbox{\hskip1pt$\sim$}}
    \raise1pt\hbox{$<$}}}         
\def\gsim{\mathrel{\rlap{\lower4pt\hbox{\hskip1pt$\sim$}}
    \raise1pt\hbox{$>$}}}         

\usepackage{graphicx}
\usepackage[figuresright]{rotating}

\newcommand{\AmS}{{\protect\the\textfont2
  A\kern-.1667em\lower.5ex\hbox{M}\kern-.125emS}}

\title{Global and Unified Analysis of Solar Neutrino Data}

\author{M.C. Gonzalez--Garcia and C. Pe\~na--Garay  
     \address{Instituto de F\'{\i}sica Corpuscular \\ 
      Universitat de  Val\`encia -- C.S.I.C\\
      Edificio Institutos de Paterna, Apt 2085, 46071 Val\`encia, 
      Spain}\thanks{Expanded version of the proceedings for the talk 
      presented at $\nu$-2000 Conference, 
      Sudbury, Canada, June 2000.  This work was supported by the 
      spanish DGICYT 
      grants PB98-0693 and PB97-1261, by the Generalitat Valenciana
      grant GV99-3-1-01 and by the TMR EU network grant ERBFMRXCT960090.}}
       
\begin{document}
\begin{abstract}
We discuss the status of the solutions of the solar 
neutrino problem in terms of oscillations of $\nu_e$ 
into active or sterile neutrinos. 
We present the results of a global fit to the full
data set corresponding to the latest Super--Kamiokande (SK) 
data on the total event 
rate, their day--night dependence and the recoil electron
energy spectrum, together with the data from Chlorine and Gallium 
experiments presented at the $\nu$-2000 conference. We show the possible 
solutions in the full parameter space for oscillations including both
MSW and vacuum, as well as quasi-vacuum oscillations (QVO) and matter effects 
for mixing angles 
in the second octant (the so called dark side). We quantify our results
in terms of allowed regions as well as the goodness of the fit (GOF) 
for the different allowed solutions. 
Our conclusion is that from the statistical point of view,  
all solutions for oscillations into active neutrinos: the large mixing
angle (LMA), the low mass (LOW), the small mixing angle (SMA) and the
QVO solutions are acceptable since they all provide a reasonable GOF 
to the full data set. The same holds for the SMA solution for oscillations 
into sterile neutrinos. LMA and LOW-QVO solutions for oscillations into 
active neutrinos seem slightly favoured over SMA solutions for oscillations 
into active or sterile neutrinos but these last two are not ruled out. 
We also analyze the dependence of these conclusions on the
uncertainty of the SSM $^8$B flux and on the removal of the data from 
one of the experimental rates.
We also present the results in the framework of four neutrino oscillations 
which allows for oscillations into a state which is a 
combination of active and sterile neutrino.
\end{abstract}

\maketitle
\section{Introduction: Experimental Status}
The sun is a source of $\nu_e's$ which are produced in the different
nuclear reactions taking place in its interior. Along this talk we will 
use the $\nu_e$ fluxes from Bahcall--Pinsonneault 
calculations~\cite{bp98} which we refer to as the solar standard model (SSM).
\begin{figure}[htbp]
\vskip -0.7cm
\begin{center}
\begin{turn}{-90}
\includegraphics[scale=0.3]{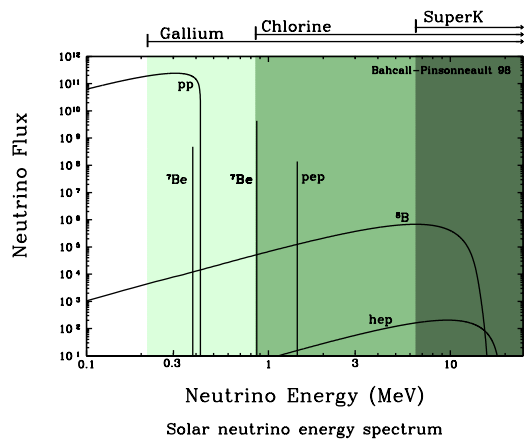}
\end{turn}
\end{center}
\vskip -0.7cm
\caption{Bahcall--Pinsonneault solar neutrino fluxes.} 
\label{fluxes}
\end{figure} 

At present these neutrinos have been detected at the Earth by seven 
experiments which use different detection techniques:\\
-- Homestake (chlorine)~\cite{chlorine}
$\nu_e + ^{37}Cl \rightarrow ^{37}Ar + e^-$ \\
-- SAGE~\cite{sage} and GALLEX+GNO~\cite{gallex,gno} (gallium)
$\nu_e + ^{71}Ga \rightarrow ^{71}Ge + e^-$\\
-- Kamiokande and SK~\cite{superk,suzuki} 
$\nu_e e¯$ scattering on water \\
Due to the different energy threshold for the detection
reactions, these experiments  are sensitive to different parts of the 
solar neutrino spectrum as represented in Fig.~\ref{fluxes}.
They all observe a deficit between 30 and 60 \% 
which seems to be energy dependent mainly due to the lower 
Chlorine rate. 
\begin{figure}[htbp]
\vskip -0.7cm
\begin{center}
\includegraphics[scale=0.5]{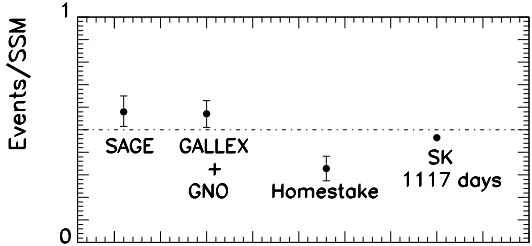}
\end{center}
\vskip -0.7cm
\caption{Measured solar neutrino event rates normalized to the SSM 
prediction.}
\label{rates}
\end{figure}

To the measurements of these six experiments we have to add also the
new results from SNO~\cite{sno} first presented in this conference.
They are however still not in the form of definite measured rates which
could be included in this analysis.
  
In this conference SK has also presented their results 
after 1117 days of data taking on:

-- The recoil electron energy spectrum: SK has measured the dependence
of the even rates on the recoil electron energy spectrum divided in
18 bins starting at 5.5 MeV. They have also reported the results
of a lower energy bin 5 MeV $<E_e<$5.5 MeV, but its systematic errors
are still under study and it is not included in their nor our analysis. 
The spectrum  shows no clear distortion with $\chi^2_{flat}=13/(17dof)$.
17 dof=18 bins-1 free normalization.
\begin{figure}[htbp]
\vskip -0.7cm
\begin{center}
\includegraphics[scale=0.45]{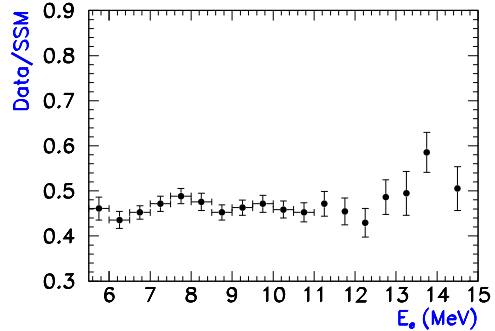}
\end{center}
\vskip -0.7cm
\caption{Recoil electron energy spectrum measured by SK normalized to the SSM 
prediction.}
\end{figure} 

--The zenith Angle Distribution (Day/Night Effect) which measures
the effect of the Earth Matter in the neutrino propagation. We have
included in the analysis the experimental results from SK  on the
zenith angle distribution of events taken on 5 night periods and the
day averaged value. SK finds few more events at night than 
during the day but the corresponding Day--Night asymmetry
\begin{equation}
A_{D/N}=\frac{D-N}{\frac{D+N}{2}}=-0.034\pm 0.022 \pm 0.013
\end{equation} 
is only 1.3$\sigma$ away from zero.
\begin{figure}[htbp]
\vskip -0.7cm
\begin{center}
\includegraphics[scale=0.45]{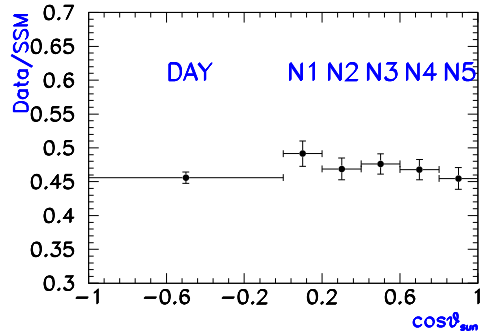}
\end{center}
\vskip -0.7cm
\caption{Zenith angle distribution measured by SK normalized to the SSM 
prediction.}
\end{figure} 

In order to combine both the Day--Night information and the
spectral data SK has also presented separately the measured recoil
energy spectrum during the day and during the night. This will be
referred in the following as the day--night spectra data which
contains $2\times 18$ data bins. 

The most generic and popular explanation of the solar neutrino anomaly is
in terms of neutrino masses and mixing leading to oscillations 
of $\nu_e$ into an active ($\nu_\mu$ and/or  $\nu_\tau$) or sterile 
neutrino, $\nu_s$. We pass now to discuss some issues that have been
raised lately in the literature concerning the computation of
the corresponding neutrino survival probabilities in the full 
range of mass and mixing relevant to the solar neutrino problem.

\section{Survival Probabilities: QVO and the Dark side}

The presence of neutrino mass and mixing imply the possibility 
of neutrino oscillations. For solar neutrinos 
we know that depending on the range of mass and mixing 
$\nu_e$ can undergo oscillations either in {\sl
vacuum}~\cite{Glashow:1987jj} or via the matter-enhanced 
{\sl MSW mechanism}~\cite{msw}. However this {\sl broken} picture of 
solar neutrino oscillations contains a set of approximations which,
as we discuss next, are not longer phenomenologically 
valid when performing the analysis of the solar neutrino data. 
In order to clarify this issue let us first review the calculation of
the solar neutrino survival probability in the two--neutrino mixing case.

The survival amplitude for a solar $\nu_e$ neutrino of
energy $E$ at a detector in the Earth can be written as:
\begin{equation}
A_{ee}=\sum_{i=1}^2 A^S_{e\,i}\,A^E_{i\,e}\,\exp[-im_i^2 (L-r)/2E]~. 
\end{equation}
Here $A^S_{e\,i}$ is the amplitude  of the transition  
$\nu_e \rightarrow \nu_i$ ($\nu_i$ is the $i$-mass eigenstate) 
from the production point to the Sun surface,  $A^E_{i\,e}$ is the 
amplitude of the transition  $\nu_i \rightarrow \nu_e$ 
from the Earth surface to the detector, and the propagation in vacuum 
from the Sun to the surface of the  Earth is given by the exponential. 
$L$ is the distance between the center of the Sun and the surface of 
the Earth, and $r$ is the distance between the neutrino production 
point and the surface of the Sun.  
The corresponding survival probability $P_{ee}$ is then given by:
\begin{equation} 
P_{ee}=P_1P_{1e}+P_2P_{2e}+2\sqrt{P_1P_2P_{1e}P_{2e}}\cos\xi
\label{Pee}
\end{equation}
Here $P_i\equiv |A^S_{e\,i}|^2$ is the probability that the solar 
neutrinos reach the surface of the Sun  as $|\nu_i\rangle$, 
while $P_{ie} \equiv  |A^E_{i\,e}|^2$ is the probability of $\nu_i$ 
arriving at the surface of the Earth to be detected as a $\nu_e$. 
Unitarity implies $P_1+P_2=1$ and $P_{1e}+P_{2e}=1$. The phase $\xi$ is 
given by 
\begin{equation} 
\xi=\frac{\Delta m^2 (L-r)}{2E}+\delta\, ,
\end{equation}
where $\delta$ contains the phases due to propagation in the
Sun and in the Earth and can be safely neglected.
In the evaluation of both $P_1$ and $P_{2e}$ the effect of coherent
forward interaction with the Sun  and Earth matter must be taken
into account.

From  Eq.~(\ref{Pee}) one can recover more familiar expressions for 
$P_{ee}$:  

(1) For $\Delta m^2/E\lsim 5 \times 10^{-17}$ eV, the matter effect 
suppresses flavour transitions  both in the Sun and the Earth. 
Consequently, the probabilities $P_1$ and $P_{2e}$ are simply
the projections of the $\nu_e$ state onto the mass eigenstates:
$P_1 = \cos^2\theta$,    $P_{2e} = \sin^2 \theta$. 
In this case we are left with the 
standard vacuum oscillation formula:
\begin{equation}
P_{ee}^{\rm vac}=1-\sin^2 2\theta \sin^2(\Delta m^2 (L-r)/4E)
\label{pvac}
\end{equation}
which describes the oscillations on the way from the surface of the Sun to 
the surface of the Earth. The probability is symmetric 
under the change of octant  
$\theta \leftrightarrow \frac{\pi}{2}-\theta$ and change of mass sign
$\Delta m^2\leftrightarrow -\Delta m^2$. Notice that the simultaneous
application of both symmetries translates simply into a relabelling
of the mass eigenstates $\nu_1\leftrightarrow\nu_2$. Therefore  
only one of these symmetries is physically independent.
This means that we can take $\Delta m^2>0$ without loss of generality 
and keep $0<\theta<\frac{\pi}{2}$. Then the symmetry of the
survival probability under the 
change of octant  $\theta \leftrightarrow \frac{\pi}{2}-\theta$ 
implies that each point in the ($\Delta m^2$, $\sin^2(2\theta)$) 
parameter space corresponds to two physically independent solutions 
one in each octant.  

\begin{figure}[htbp]
\vskip -0.7 cm
\begin{center}
\includegraphics[scale=0.45]{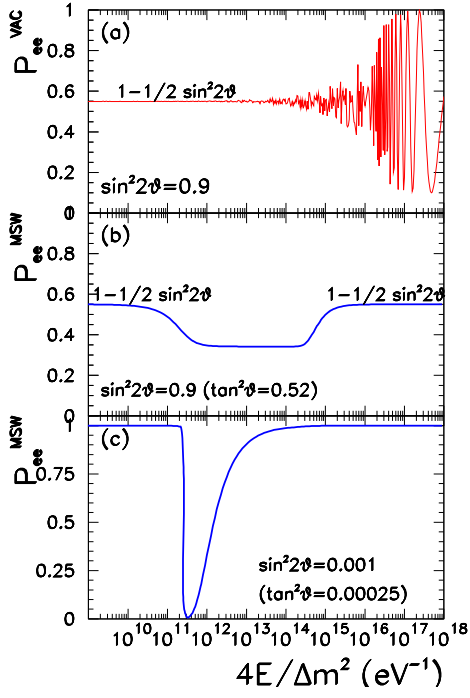}
\end{center}
\vskip -0.7 cm
\caption{$P_{ee}$ as a function of $4E/\Delta{m}^2$.}
\label{fig:probs}
\end{figure}

Averaging Eq.(\ref{pvac}) over the Earth Orbit $L(t)=L_0 [ 1 - 
\varepsilon\cos 2\pi \frac{t}{T}]$ one gets :
\begin{equation}
\begin{array}{lll} 
\langle P^{vac}_{ee}\rangle 
& = & 1 - \frac{1}{2}\sin^2{2\theta} \\ 
& & \left[ 1 - \cos\!\left(\frac{{\Delta{m}^2}{L_0}}
{ 2E}\right) 
{ J_{0}}\!\left( 
A=\frac{{ \varepsilon \Delta{m}^2}{L_0}}{{2E}}
\right)\right] 
\end{array}
\end{equation} 
where $\varepsilon=0.0167$ is the orbit eccentricity and $J_0$ is the
Bessel function. In Fig. \ref{fig:probs}.a we display the value of 
$\langle P^{vac}_{ee}\rangle$ as a function of 
$4E/\Delta{m}^2$.
As seen in the figure for large values of $\Delta m^2$ the probability
averages out to a constant value $1-\frac{1}{2}\sin^2(2\theta)$.

\begin{figure}[htbp]
\vskip -0.7cm
\begin{center}
\includegraphics[scale=0.5]{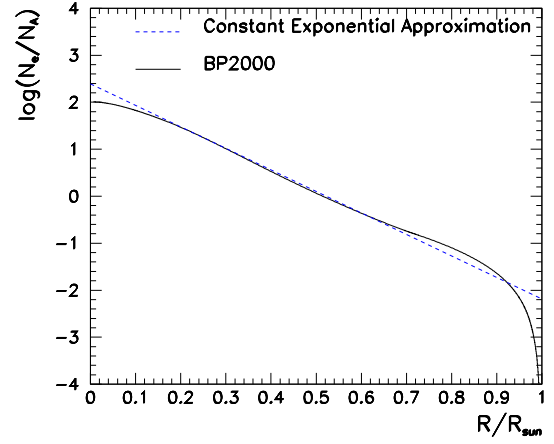}
\end{center}
\vskip -0.7cm
\caption{Solar density profile for BP2000 model.} 
\label{rho}
\end{figure}

(2) For $\Delta m^2/E\gsim 10^{-14}$ eV, the last term in Eq.~(\ref{Pee})
vanishes and we recover the incoherent MSW survival probability.
In this case $P_1$ and $P_{2e}$ must be obtained by solving the
evolution equation of the neutrino states in the Sun and the Earth
matter respectively:
\begin{eqnarray}
-i \frac{d}{dt}
\left(\hspace*{-0.2cm}
\begin{array}{c}{ \nu_e}\\{ \nu_X} 
\end{array}\hspace*{-0.2cm}\right)
= \nonumber
\\ 
\left(\hspace*{-0.2cm}\begin{array}{ll} 
{ V_e}+\frac{\Delta m^2}{2 E} {\cos 2\theta} &
-\frac{ \Delta m^2}{2 E} { \sin 2\theta} \\
-\frac{ \Delta m^2}{2 E} { \sin 2\theta} &
{ V_X}-\frac{ \Delta m^2}{2 E} { \cos 2\theta} 
\end{array}\hspace*{-0.2cm}\right)
\left(\hspace*{-0.2cm}\begin{array}{c}
{ \nu_e}\\ { \nu_X} \end{array}\hspace*{-0.2cm}\right) 
\end{eqnarray}
with
\begin{eqnarray}
{ V_e} = \frac{\sqrt{2}G_F} {M} ({ N_e-\frac{1}{2}N_n})  
      &    & { V_s}= 0\nonumber  \\
{ V_\mu}={ V_\tau} \frac{\sqrt{2}G_F}{M} 
({ -\frac{1}{2}N_n}) & & 
\label{potentials}
\end{eqnarray}
where $N_{e(n)}(r)$ 
is the electron (nucleon) number density which
are proportional to the Sun or Earth matter density. In Fig.~\ref{rho}
we show $N_{e}(r)$ for the Sun. As seen in the figure the Sun 
density profile for $r< 0.9 R_\odot$ 
can be very well approximated by an exponential 
$N(r)= N_{0} \exp(-r/r0)$ with constant exponent slope $r_0=R_\odot/10.54$.

The approximate solution for the evolution in the Sun takes the 
well-known form 
\begin{equation}
P_1= \frac{1}{2} + (\frac{1}{2} - P_{c})\cos(2\theta_{m,0})
\label{P1}
\end{equation}
where $P_{c}$ denotes the standard Landau-Zener probability
\cite{LZ} and $\theta_{m,0}$ is the mixing angle in matter
at the neutrino production point:
\begin{eqnarray} 
\cos(2\theta_{m,0})& = & 
\frac {\Delta{m}^2 c2 - A_0} 
{\sqrt{ (\Delta{m}^2 c2 - A_0)^2 
+(\Delta{m}^2 s2)^2 }} \label{thetam} 
\\
P_{c}&=&\frac{\exp[-\gamma \sin^2\theta]-\exp[-\gamma]}{1-\exp [- \gamma]} 
\label{PLZ}   
\\
\gamma &= & \pi \frac{\Delta{m}^2}{E}
\left[\left|\frac{d\ln {
N_e(r)}}{dr}\right|_{r=r_{res}}\right]^{-1}
\nonumber
\end{eqnarray}
with $A_0= 2E({V_e-V_X})$ evaluated at the production point, 
$c2=\cos(2\theta)$ and $s2=\sin(2\theta)$.  
For the approximation of exponential density profile 
$\gamma= \pi \frac{\Delta{m}^2}{E}r_0$ which is independent 
of the point in the Sun where the resonance takes place. 
Improvement over this ``constant slope'' exponential density profile 
approximation can be obtained by numerically deriving the exact 
$N_e(r)$ profile at the resonant point. In this case  
$\gamma= \pi \frac{\Delta{m}^2}{E}r_0(r_{res})$. 
 
The physical interpretation behind Eq.(\ref{P1}) is very simple.
At the  production point $\nu_e$ has a projection over the 
mass eigenstate $\nu_1$ given by the mixing angle in matter $\theta_{m,0}$.
This neutrino evolves adiabatically till the resonant point where 
both mass eigenvalues become closer and the neutrino has a probability
$P_c$ of "jumping" into the other mass eigenstate (or staying in the
state $\nu_1$  with a probability 1-$P_c$).
During the day no Earth matter effect is to be 
included and the survival probabilities $P_{ie}$ are obtained 
by simple projection of the   
$\nu_e$ state onto the mass eigenstates $P_{2e,DAY} =
1-P_{1e,DAY}=\sin^2\theta$ and one obtains
\begin{equation}
P^{MSW}_{ee,DAY}= \frac{1}{2} + (\frac{1}{2} - 
P_{c})\cos(2\theta_{m,0})\cos(2\theta)
\label{pmsw}
\end{equation}
In Fig.~\ref{fig:probs}.b (~\ref{fig:probs}.c) we plot the 
this survival probability as a function of $4E/\Delta m^2$ for
large (small) mixing angle.
  
Let us make some remarks concerning Eq.(\ref{pmsw}):

$(i)$ In both limits of very large and very small $E/\Delta m^2$ 
$P^{MSW}_{ee}\rightarrow 1-\frac{1}{2}\sin^2(2\theta)$ 
(See Fig.\ref{fig:probs}.b). For large
$E/\Delta m^2$, $P_c=\cos^2\theta$ and $\cos(2\theta_{m,0})=-1$. 
While for small $E/\Delta m^2$, $P_c=0$ and 
$\cos(2\theta_{m,0})=\cos(2\theta)$. So in principle
one could expect a perfect connection between the 
asymptotic small $E/m^2$ probability for the vacuum oscillation
regime and the large $E/m^2$ behaviour of the MSW regime. However as 
seen when comparing panels (a) and (b) of Fig.\ref{fig:probs} 
there is an intermediate range,  
$ 2 \times 10^{14} \lsim  4 E/\Delta m^2  \lsim 10^{16}$ eV, 
where both adiabaticity is violated and the $\cos\xi$ coherent term should be 
taken into account. The result is similar to vacuum oscillations 
but with small matter corrections. We define this case as QVO  
\cite{panta,fried,threev}. The range of $E/m^2$  
for the QVO regime depends on the value of $E/m^2$ 
for which the MSW probability in Eq.~\ref{pmsw} 
acquires the asymptotic value 
$1-\frac{1}{2}\sin^2(2\theta)$, the smaller $E/m^2$ the more separated
the MSW and vacuum regimes are, and the narrower the QVO region is. 

$(ii)$ Due to matter effects $P^{MSW}_{ee}$ 
is only symmetric under simultaneous 
$(\Delta m^2, \theta) \rightarrow  
(-\Delta m^2, \frac{\pi}{2}-\theta)$. 
For ${\Delta m^2>0}$ the resonance is only possible for 
$\theta <\frac{\pi}{4}$ and  MSW solutions are usually plotted in the 
$(\Delta m^2, \sin^2(2 \theta))$  plane assuming that now each point
on this plane represents only one physical solution with $\theta$ in
the first octant.
But in principle non-resonant solutions are also possible for 
$\theta >\frac{\pi}{4}$, the so called {\sl dark side}
\cite{dGFM,GGPG,three,four}. 

It is clear from these considerations that in order to compute the
survival probability for solar neutrinos, valid for any value of
the neutrino mass and mixing, the full expression (\ref{Pee}) has
to be evaluated. The results presented in the following sections have 
been obtained using
the general expression for the survival probability in Eq.~(\ref{Pee})
with $P_1$ and $P_{2e}$ found by numerically solving the evolution
equation 
in the Sun and the Earth matter. For $P_1$ we use the electron number 
density of 
BP2000 model \cite{BP00}. For $P_{2e}$ we integrate numerically the evolution 
equation in the Earth matter using the Earth density profile given in the 
Preliminary Reference Earth Model (PREM) \cite{PREM}.

\begin{figure}[htbp]
\vskip -0.7cm
\begin{center}
\includegraphics[scale=0.4]{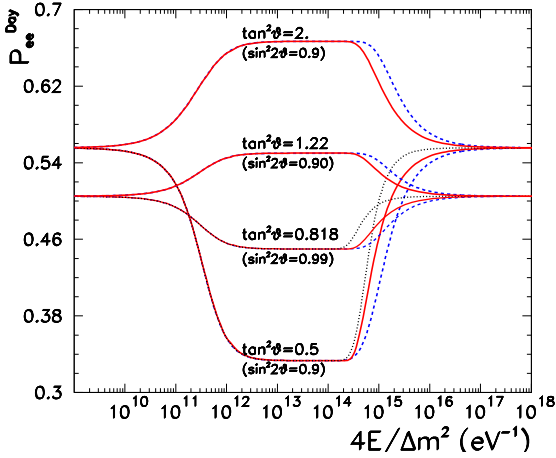}
\end{center}
\vskip -0.7cm
\caption{Survival probabilities as a function of $4E/\Delta m^2$ for 
different values of the mixing angle. The full line corresponds to the
numerical solution while the dashed (dotted) lines corresponds to the
analytical approximation for exponential density profile with constant
(point dependent) slope.} 
\label{probs}
\end{figure}

Before moving to the next section we want to discuss the validity
of the analytical and semi-analytical approximations based on the exponential 
profile for the sun density in the
evaluation of the survival probability when used to describe the 
matter effects in the QVO regime as well as for mixing angles in 
the second octant. In order to illustrate this point we display in
Fig.~\ref{probs} the survival probability $P_{ee}$ (without the oscillating
$\cos\xi$ term) 
as a function of $4E/\Delta m^2$ for different values of the mixing angle 
obtained by our numerical solution as well as from the
corresponding analytical approximations (\ref{pmsw}). 
We discuss two analytical approximations: the exponential approximation
with constant slope $r_0=R_\odot/10 54$ and with point
dependent slope $r_0(r)$. 

We see from Fig.~\ref{probs} that the analytical results 
differ from the results of the numerical calculations. In particular, 
the analytical result with constant exponential slope 
shows a ``slower'' transition to the vacuum oscillation regime or, in other
words, it overestimates the size of the matter effects in the 
QVO region. The same value of the survival
probability appears for about twice larger  $E/\Delta m^2$. 
This is due to the fact that for such values of $E/\Delta m^2$  
the adiabaticity breaking occurs very close to the Sun edge where the
density falls much faster than the exponential approximation as
shown in Fig.~\ref{rho}.  
Similar conclusions have been drawn in Refs. \cite{fried,threev,maxmix}.
On the other hand the analytical approximation with point dependent
slope shows  a ``faster'' transition to the vacuum oscillation regime or, 
in other words, it underestimates the size of the matter effects in the 
QVO region. 

Notice also that, originally, Eq.~(\ref{PLZ}) 
was derived for a mixing angle $\theta<\frac{\pi}{4}$ where resonant 
enhancement is possible. However for the constant slope exponential profile, 
$\gamma$ is constant
(independent of the resonant condition) and both Eq.~(\ref{thetam}) 
and Eq.~(\ref{PLZ}) can be analytically continued into the second octant, 
$\theta > \pi/4$, and used to compute the corresponding survival 
probability. This is illustrated in 
Fig.~\ref{probs}. As seen in the figure, for values of the mixing angle
close to maximal mixing $\theta=\pi/4$, the survival probability is 
mirror-symmetric in the first and second octant, while, as expected, 
the symmetry breaks down as we depart from maximal mixing. 
We also see that the analytical approximation overestimates 
the size of matter effects in the QVO region on the second octant as well. 
     
\begin{figure}[htbp]
\vskip -0.7cm
\begin{center}
\includegraphics[scale=0.4]{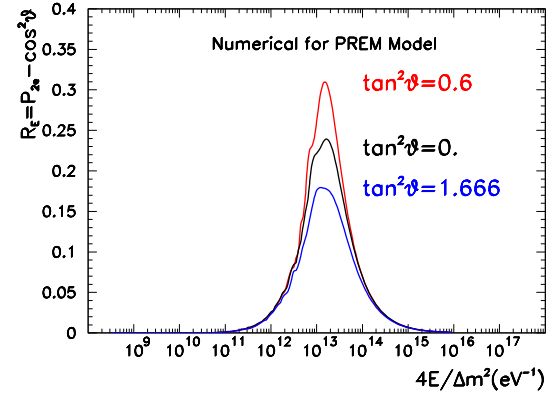}
\end{center}
\vskip -0.7cm
\caption{Regeneration of $\nu_e$ in the Earth matter 
as a function of the $4E/\Delta m^2$ for different mixing angles.}
\label{pe}
\end{figure}

Finally we want to comment briefly on the Earth matter effects. 
In Fig.~\ref{pe} 
we show the Earth regeneration factor $R_E=P_{2e,NIGHT}-P_{2e,DAY}$ 
(obtained numerically) 
as a function of $4E/\Delta m^2$ for two values of the  mixing angle in
the first and second octant. We see that:

(a) Earth Matter effects are only relevant in the ``pure'' MSW regime 
$4E/\Delta m^2\lsim 10^{-14}$ eV$^{-1}$, while they are very small for  
QVO. 

(b) Earth Regeneration is always positive (it always leads to an increase
of the $\nu$ flux at night) for mixing angles both 
in the first or in the second octant. 
\section{Two--neutrino Oscillations}

We now describe the results of the analysis of the solar neutrino 
data in terms of $\nu_e$ oscillations into active or sterile neutrinos.
For details on the statistical analysis applied to the different
observables we refer to Ref.~\cite{gonzalez}. 

We first determine the allowed range of oscillation parameters using
only the total event rates of the Chlorine, Gallium and
SK experiments shown in Fig.\ref{rates}.
For the Gallium experiments we have used the weighted average of the results
from GALLEX+GNO and SAGE detectors and we do not include
the Kamiokande data as it is well in
agreement with the results from the SK experiment and
the precision of this last one is much higher. 

Using the predicted fluxes from the BP98 model the $\chi^2$ for the
total event rates is $\chi^2_{SSM}=60$ for 3 d.~o.~f.
This means that the SSM together with the SM of particle interactions
can explain the observed data with a probability lower than
$10^{-12}$. 

\begin{figure*}[htbp]
\vskip -0.5cm
\begin{center}
\includegraphics[scale=1.3]{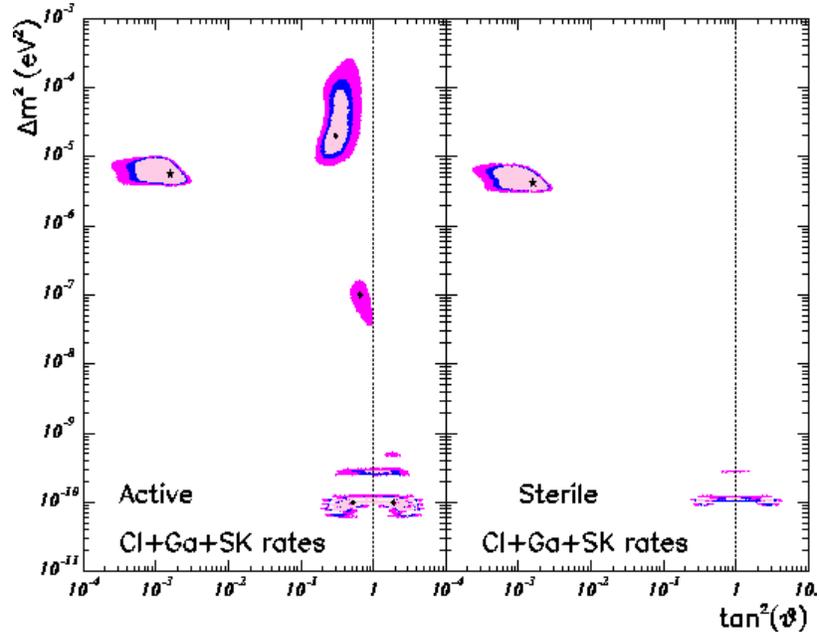}
\end{center}
\vskip -0.9cm
\caption{90, 95 and 99 \% CL allowed regions from the analysis of the total 
rates only. The global minimum is marked with a star while the local minima
are denoted with a dot.}
\label{2fr}
\end{figure*}

The allowed regions in the oscillation parameter space are shown in 
Fig.~\ref{2fr}. 
In the case of active--active neutrino oscillations we 
find that the best--fit point is obtained for the 
SMA solution which has a probability or goodness of the fit 
of 50 
the LMA and LOW solutions.
In the region of vacuum oscillations we find a ``tower'' of regions
around local minima corresponding to the oscillation wavelength being an 
entire fraction of the Sun-Earth distance. In Table ~\ref{minima} we give
the values of the parameters in these minima as well as the GOF corresponding
to each solution. 

Notice that following the standard procedure,  
the allowed regions for a given set of observables 
are defined in terms of shifts of the
$\chi^2$ function for those observables {\sl with respect to the global
minimum in the plane}. Defined this way, the size of a region depends on the
{\sl relative} quality of its local minimum with respect to the global minimum
but from the size of the region we cannot infer the actual {\sl absolute} 
quality of the description in each region. In order to give this information 
we list in Table ~\ref{minima} the GOF for each solution obtained from the
value of $\chi^2$ at the different minima.  
We see that for oscillations into active neutrinos the best solution by 
far for the description of the rates is the SMA. 

For oscillations into 
sterile neutrinos we find only the SMA solution and some small region 
for LOW-QVO. The LMA and most of the LOW solutions are not acceptable
for oscillation into sterile neutrinos. 
Unlike active neutrinos which lead to events in the
SK detector by interacting via neutral current (NC) with the
electrons, sterile neutrinos do not contribute to the SK
event rates.  Therefore a larger survival probability for $^8B$
neutrinos is needed to accommodate the measured rate. As a consequence
a larger contribution from $^8B$ neutrinos to the Chlorine and Gallium
experiments is expected, so that the small measured rate in Chlorine
can only be accommodated if no $^7Be$ neutrinos are present in the
flux. This is only possible in the SMA solution region, since in the
LMA and LOW regions the suppression of $^7Be$ neutrinos is not enough.
Notice also the SMA region for oscillations into sterile neutrinos is 
slightly shifted downwards as compared with the active case. This is due
to the small modification in the neutrino survival probability induced
by  the different matter potentials. The matter potential for sterile
neutrinos is smaller than for active neutrinos due to the negative NC
contribution proportional to the neutron abundance 
(see Eq.~(\ref{potentials})). 
For this reason the resonant condition for sterile neutrinos is achieved at
lower $\Delta m^2$.  

On the other hand SK data on the day--night variation of the event rates 
and the spectrum data lead to excluded regions in the parameter space. 
Since both the observed zenith angular dependence and the recoiled
energy spectrum are compatible with the no--oscillation hypothesis 
(up to an overall normalization which is taken free), these
measurements translate in the exclusion of those regions of the oscillation
parameter space where the expectation for those observables is far different
from the no--oscillation one. 
In this way we see in Figs.~\ref{2fz},~\ref{2fs} and~\ref{2fsdn} the
regions excluded by the zenith angle, average spectrum and combined
day--night spectra data. The zenith angular data excludes the region
where large regeneration in the Earth is expected  (See Fig.~{\ref{pe}) 
which for neutrino energies of few MeV (above SK threshold) occurs for
few $10^{-8}\lsim \Delta m^2$/eV$^2\lsim 10^{-5}$. In Table ~\ref{minima}
we show the effect of the inclusion of this observation in the position
of the different local minima. We see that when combined with the 
zenith angle data the LMA minimum is shifted up and the LOW minimum is
shifted down towards values of parameters where the Earth regeneration 
is smaller. 
\begin{figure*}[htbp]
\vskip -0.5cm
\begin{center}
\includegraphics[scale=1.3]{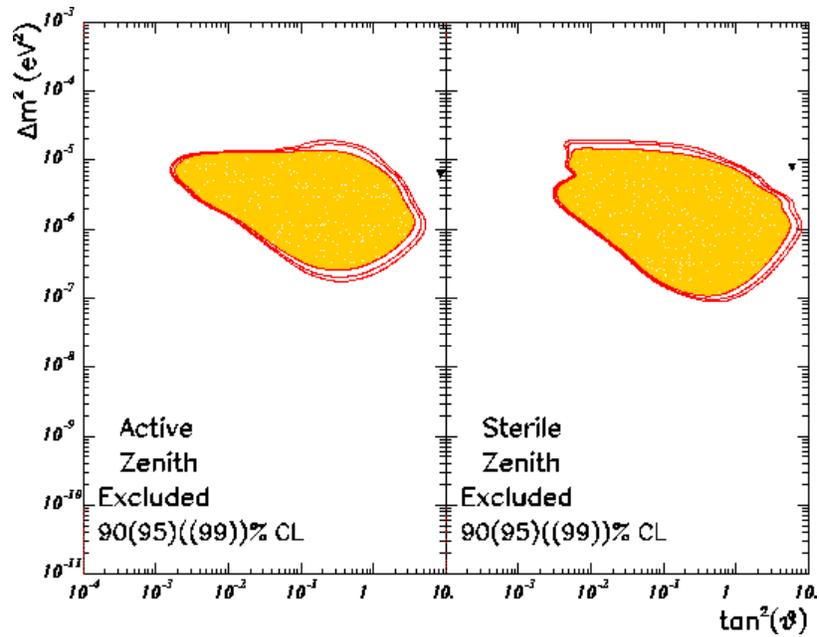}
\end{center}
\vskip -0.9cm
\caption{99 (shadow), 95 and 90 excluded regions from the analysis of 
SK measured zenith angle dependence. The inverted triangle
represents the minimum used to define the regions.}
\label{2fz}
\end{figure*}

The excluded region from the non observation of any clear
distortion of the energy spectrum is shown in Fig.~\ref{2fs}. 
Its main effect is to suppress the vacuum oscillation solution where a
large distortion of the spectrum is expected due to the ``imprints'' 
of the oscillation wavelength. On the other hand for the SMA 
a positive tilt of the spectrum is also predicted due to the raising 
with the neutrino energy of the survival probability 
(see Fig.~\ref{fig:probs}.c). For this reason the local SMA minimum shifts
to lower values of the mixing angle where a flatter probability is
expected. On the other hand both LMA and LOW solutions predict a rather
flat spectrum and in consequence the position of these minima 
are little affected. 
\begin{figure*}[htbp]
\vskip -0.5cm
\begin{center}
\includegraphics[scale=1.3]{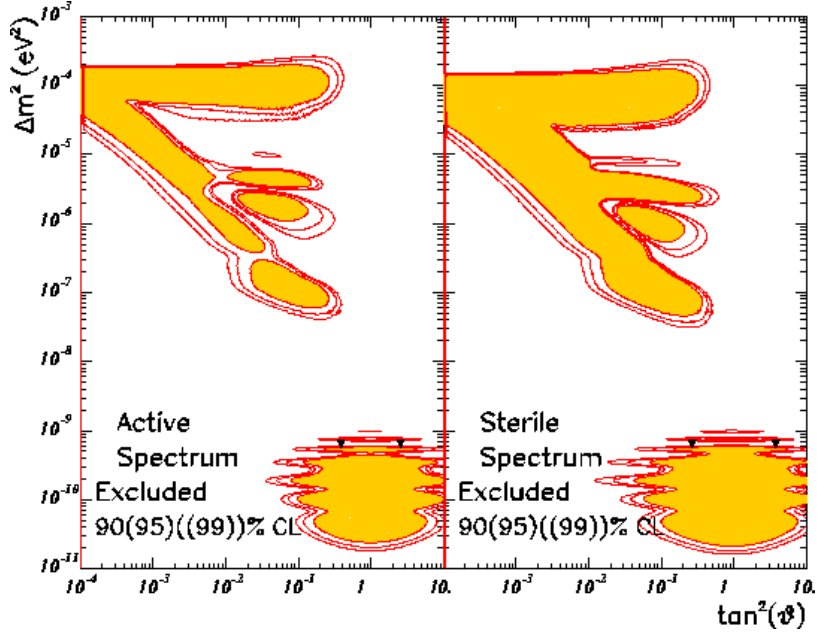}
\end{center}
\vskip -0.9cm
\caption{99 (shadow), 95 and 90 excluded regions from the analysis of 
the dally averaged SK measured spectrum. The inverted triangle
represents the minimum used to define the regions.}
\label{2fs}
\end{figure*}

Finally Fig.~\ref{2fsdn} shows the excluded region from the SK day--night
spectra which contains simultaneously the information on the Earth
regeneration effect and the energy dependence of the survival probability.
Combining statistically this information with the data from the total 
rates (what implies 3+2x18 data points with 1 free normalization for
the spectra) we obtain the allowed regions which we display in 
Fig.~\ref{global} while the corresponding GOF for the global 
solutions are given in 
Table~\ref{minima}. 
There are some points concerning these results that
we would like to stress:

(a) Despite giving a worse fit to the observed total rates, once the 
day--night spectra data is included the LMA gives the best 
fit. This is mainly driven by the flatness of the spectrum and it was
already the case with the last year data as pointed out in 
Ref.~\cite{gonzalez}.

(b) The GOF of the LOW solution has increased considerably as it describes
the spectrum data very well despite it gives a bad fit to the
global rates. LOW and QVO regions are connected at the 99 \%CL and they 
extend into the second octant so maximal mixing is allowed at 99 \% CL 
for $\Delta m^2$ in the LOW-QVO region. Notice that part of this 
allowed region would be missing when making the {\sl old fashioned} 
analysis in terms of pure  MSW or vacuum oscillations.

(c) What is the situation for the SMA solution?.  
Superimposing the excluded region from the day--night spectra in 
Fig.~\ref{2fsdn} with the allowed region from the global rates in 
Fig.~\ref{2fr} we see that almost  the full allowed region at 95 \%CL lays
below the 95\% CL excluded region from the day--night spectra. 
However this does not mean that the SMA solution is ruled out at
that CL. The result from the correct statistically combined analysis 
is what we show in Fig.~\ref{global} and more important in 
Table ~\ref{minima}. From this results we learn that the SMA can describe
the full data set with a probability of 34\%,  but it is now shifted
to smaller mixing angles to account for the flatter spectrum.  

(d) Similar statement as (c) holds for the SMA solution for sterile neutrinos.

Thus our conclusion is that from the statistical point of view all 
solutions are acceptable since they all provide a reasonable GOF to the
full data set. 
LMA and LOW-QVO solutions for oscillations into active neutrino seem slightly 
favoured over SMA solutions for oscillations into active or sterile neutrinos 
but these last two are not ruled out. 

Let's comment now how these conclusions may depend 
on the specific features of the SSM model and/or the 
analyzed data. In order to do so we are going to discuss two 
departures from the previous global analysis:

(a) effect of the uncertainty in the predicted SSM boron flux, and 

(b) effect of removing from the analyzed data one of the measured rates.

Concerning (a) in Fig.~\ref{borofree} we show the results of the
analysis for active and sterile oscillations when we allow the departure
from SSM normalization for the $^8$B flux. In doing so
we have treated the boron flux as a free parameter that must be fixed by 
the experiments, in particular by SK. The quality of the different minima is 
shown in the last line of Table \ref{minima}. Comparing with the 
results for fixed SSM $^8B$ 
flux we see that allowing free $^8B$ 
normalization leads to both an 
improvement of the quality of the SMA and LMA solutions but the effect is
more important for the SMA. The small improvement of the LMA solution is due
to the fact that allowing for a free $^8B$ flux 
leads to a better 
simultaneous agreement with Cl and SK measured total rates.
The most important effect on the SMA solution arises from the fact that
in order to account for a flatter spectrum the SMA solution has shifted to
smaller angles, what also implies a larger $^8B$ flux 
contribution to both Cl and SK. This can be compensated by the 
allowance of a lower predicted $^8B$ flux. Notice also that the LOW 
solution and QVO solutions which have the best fit very close to 
the SSM $^8B$ normalization, the quality of the fit has decreased simply 
due to the effect of the additional free parameter. 

For (b) the situation is more involved. In Fig.~\ref{two} we show the analysis 
for active and sterile oscillations where we have removed the number 
of events measured by each experiment, SK, Cl and Ga respectively, from the 
global analysis but always keeping the SK day-night spectra.
The quality of the local minima is showed in  
Table \ref{minima_two}. From the figure and the table we can draw two main 
conclusions:\\
-- The results for oscillations into active and sterile neutrinos are only 
significantly different for the LMA and LOW-QVO solutions when both SK and 
Cl rates are included in the analysis. This is consequence of the 
larger survival probability for $^8B$ neutrinos needed to accommodate 
the measured rate at SK for oscillations into sterile neutrino (due to the
absence of the NC contribution) which for large mixing angles lead to
a too large contribution to Chlorine rate. On the contrary, if the chlorine 
experiment is removed from the global analysis both oscillations into active
or sterile neutrinos provide equivalently good fits to the data and in 
particular a large region in both sides of maximal mixing is 
allowed~\cite{maxmix} .\\
-- The SMA region is disfavoured in the global analysis due to the spectrum 
of SK and this result holds  independently of the removal of one the 
measured rates. It is clear by comparing the active SMA and LMA 
columns of Table~\ref{minima_two} that the relative quality of SMA 
versus LMA worsens when SK or Cl total rates are removed from the
analysis. However once both are included (third row), the difference between
SMA and LMA descriptions decreases because, as explained before, the 
worse SMA fit to the SK spectrum is compensated by the better fit to these
two rates simultaneously. As a matter of fact, in this case the best solution
lies in the LOW region where both rates can be better fitted than for LMA 
and the spectrum data is better described than for SMA.
\begin{figure*}[htbp]
\vskip -0.5cm
\begin{center}
\includegraphics[scale=1.3]{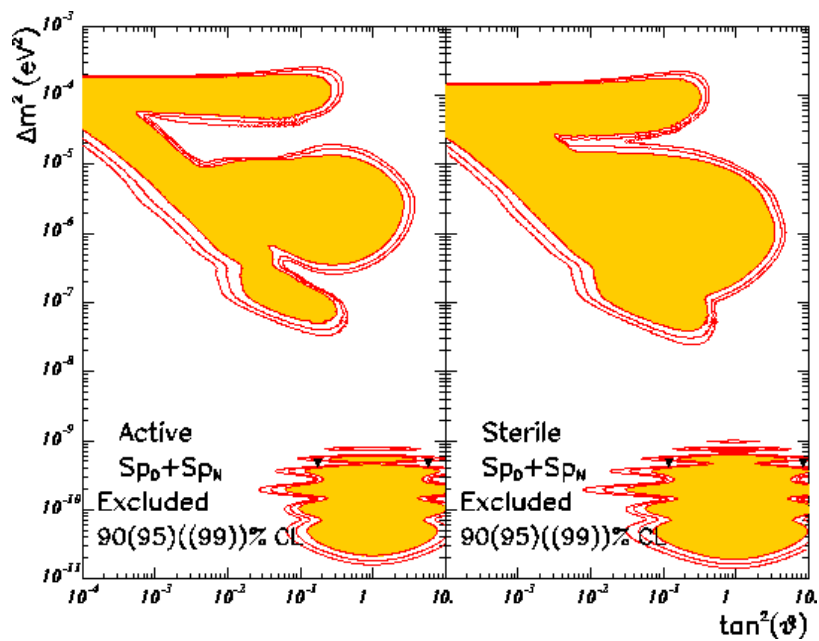}
\end{center}
\vskip -0.9cm
\caption{99 (shadow), 95 and 90 excluded regions from the analysis of 
the SK measured spectrum at night and at day. The inverted triangle
represents the minimum used to define the regions.}
\label{2fsdn}
\end{figure*}
\begin{figure*}[htbp]
\vskip -0.5cm
\begin{center}
\includegraphics[scale=1.3]{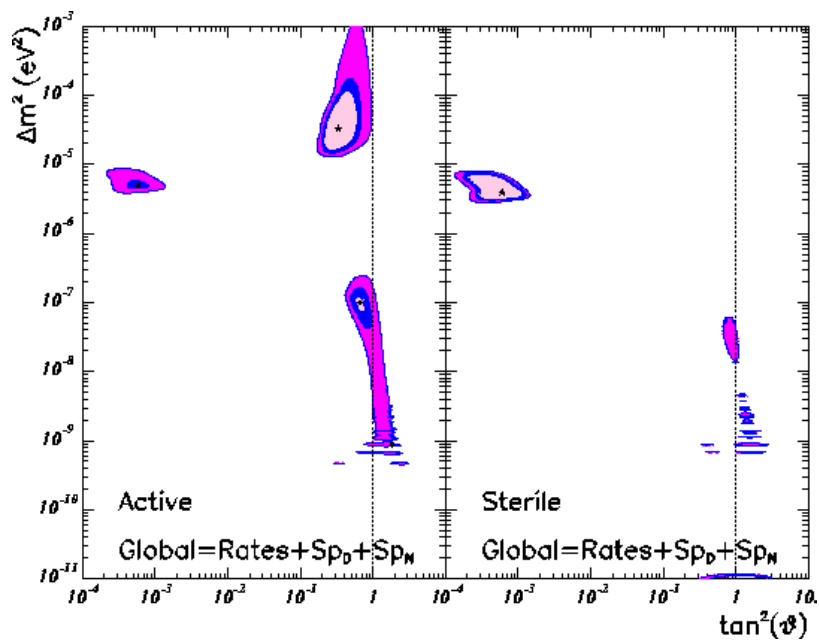}
\end{center}
\vskip -0.9cm
\caption{90, 95 and 99 \% CL allowed regions from the global analysis of  
solar neutrino data  including the total measured rates and the SK measured
spectrum at day and night. The global minimum is marked with a star while 
the local minima are denoted with a dot.}
\label{global}
\end{figure*}
\begin{table*}
\caption{Best fit points and GOF for the allowed solutions for different
combinations of observables.}
\label{minima}
\begin{center}
\begin{tabular}{|c|c|c|c|c|c|c|}
\hline
Observable &  & \multicolumn{4}{c|}{Active}  & Sterile
\\\hline            
            & & SMA & LMA & LOW & VAC-QVO & SMA  \\\cline{2-7} 
   Rates    & $\Delta m^2$/eV$^2$ & $5.5\times 10^{-6} $  
            & $1.9\times 10^{-5} $  
            & $9.2\times 10^{-8} $  
            & $9.7\times 10^{-11} $ 
            & $4.1\times 10^{-6} $     
                  \\ 
            & $\tan^2\theta$ & 0.0015     
            & 0.3   & 0.65    & 0.51 (1.94) & 0.0015  \\ \cline{2-7}             
            & Prob (\%)&  {50} \%  &  {8} \%  
            &  0.5 \% &  {2} \% &19 \%  \\\hline      

   Rates   & $\Delta m^2$/eV$^2$ & $5.6\times 10^{-6} $  
            & $5.2\times 10^{-5} $  
            & $7.9\times 10^{-8} $  
            & $9.7\times 10^{-11} $ 
            & $4.1\times 10^{-6} $     
               \\ 
+Zenith           
            & $\tan^2\theta$ & 0.0012     
            & 0.35   & 0.69    & 0.51 (1.94) &0.0015  \\ \hline             

   Rates   &  $\Delta m^2$/eV$^2$ & $5.0\times 10^{-6} $  
            & $2.1\times 10^{-5} $  
            & $9.\times 10^{-8} $  
            & $8.4\times 10^{-10} $ 
            & $3.9\times 10^{-6} $   
 \\ 
+Spectrum           
            & $\tan^2\theta$ & 0.00075     
            & 0.32   & 0.67    & 1.7 (QVO) & 0.00069  \\ \hline
    Rates   &  $\Delta m^2$/eV$^2$ & $5.0\times 10^{-6} $  
            & $3.2\times 10^{-5} $  
            & $1.\times 10^{-7} $  
            & $8.6\times 10^{-10} $
            & $3.9\times 10^{-6} $   \\ 
+Spec$_D$+Spec$_N$           
            & $\tan^2\theta$ & 0.00058     
            & 0.33   & 0.67    & 1.5 (QVO) &0.0006 \\ \cline{2-7}             
            & Prob (\%)&  {34} \%  &  {59} \%  & {40} \% &  29 \%  &30\%   
\\ \hline
    Rates   &  $\Delta m^2$/eV$^2$ & $4.9\times 10^{-6} $  
            & $3.0\times 10^{-5} $  
            & $9.7\times 10^{-8} $  
            & $8.4\times 10^{-10} $
            & $3.9\times 10^{-6} $   \\ 
+Spec$_D$+Spec$_N$           
            & $\tan^2\theta$ & 0.00046     
            & 0.26   & 0.67    & 1.7 (QVO) &0.00053 \\ \cline{2-7}             
free $^8B$  
            & Prob (\%)&  {51} \%  &  {65} \%  & {36} \% &  27 \%  &42\%  
\\\hline      
\end{tabular}
\end{center}
\end{table*}
\begin{figure*}[htbp]
\vskip -0.5cm
\begin{center}
\includegraphics[scale=1.3]{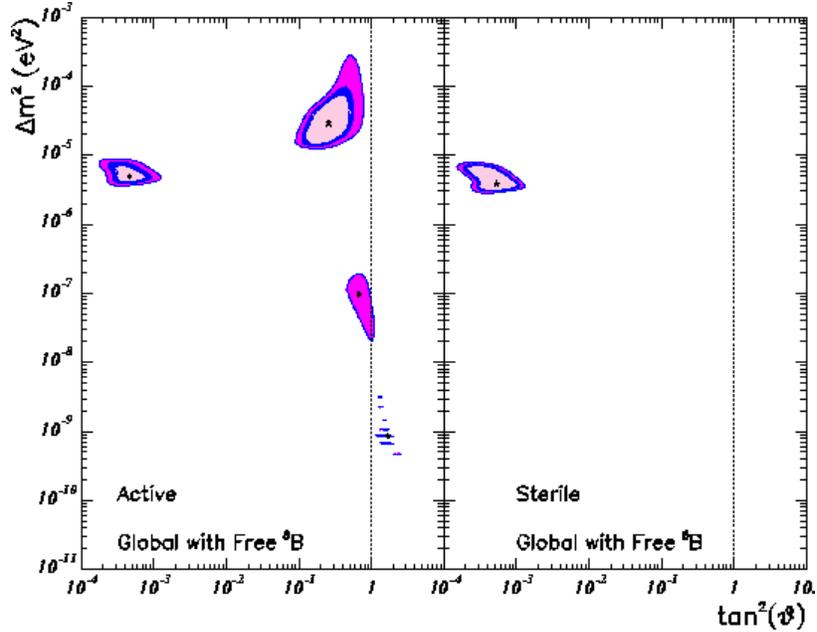}
\end{center}
\vskip -0.9cm
\caption{90, 95 and 99 \% CL allowed regions from the global analysis of  
solar neutrino data taking the boron flux as a free parameter. The global 
minimum is marked with a star while the local minima are denoted with a dot.}
\label{borofree}
\end{figure*}
\begin{table*}
\caption{$\chi^2_{min}$ and GOF in the regions of SMA, LMA 
and LOW for the global analysis when one of the fluxes is removed from the 
data (35 dof).}
\label{minima_two}
\begin{center}
\begin{tabular}{|c|c|c|c|c|c|c|c|}
\hline
Observable &  & \multicolumn{3}{c|}{Active}  & \multicolumn{3}{c|}{Sterile}
\\\hline            
            & & SMA & LMA & LOW & SMA & LMA & LOW \\\cline{2-8} 
    Rates (Ga + Cl)   
         & $\chi_{min}$ & 38.9 & 29.8 & 34.3 & 39.2 & 29.6 & 34.3\\
+Spec$_D$+Spec$_N$           
         & Prob (\%)&  {30} \%  &  {72} \%  & {50} \% & {29} \%  &  {73} \%  & {50} \%  
            \\\hline      
    Rates (Ga + SK)   
         & $\chi_{min}$ & 38.1 & 30.5 & 30.4 & 37.2 & 31.4 & 29.4\\
+Spec$_D$+Spec$_N$           
         & Prob (\%)&  {33} \%  &  {69} \%  & {69} \% & {37} \%  &  {64} \%  & {73} \%
            \\\hline      
    Rates (Cl + SK)   
         & $\chi_{min}$ & 38.0 & 33.0 & 31.3 & 39.7 & 45.7 & 37.8\\
+Spec$_D$+Spec$_N$           
         & Prob (\%)&  {33} \%  &  {56} \%  & {65} \% &  {27} \%  &  {12} \%  & {34} \%
            \\\hline      
\end{tabular}
\end{center}
\end{table*}
\begin{figure*}[htbp]
\vskip -0.5cm
\begin{center}
\includegraphics[scale=1.3]{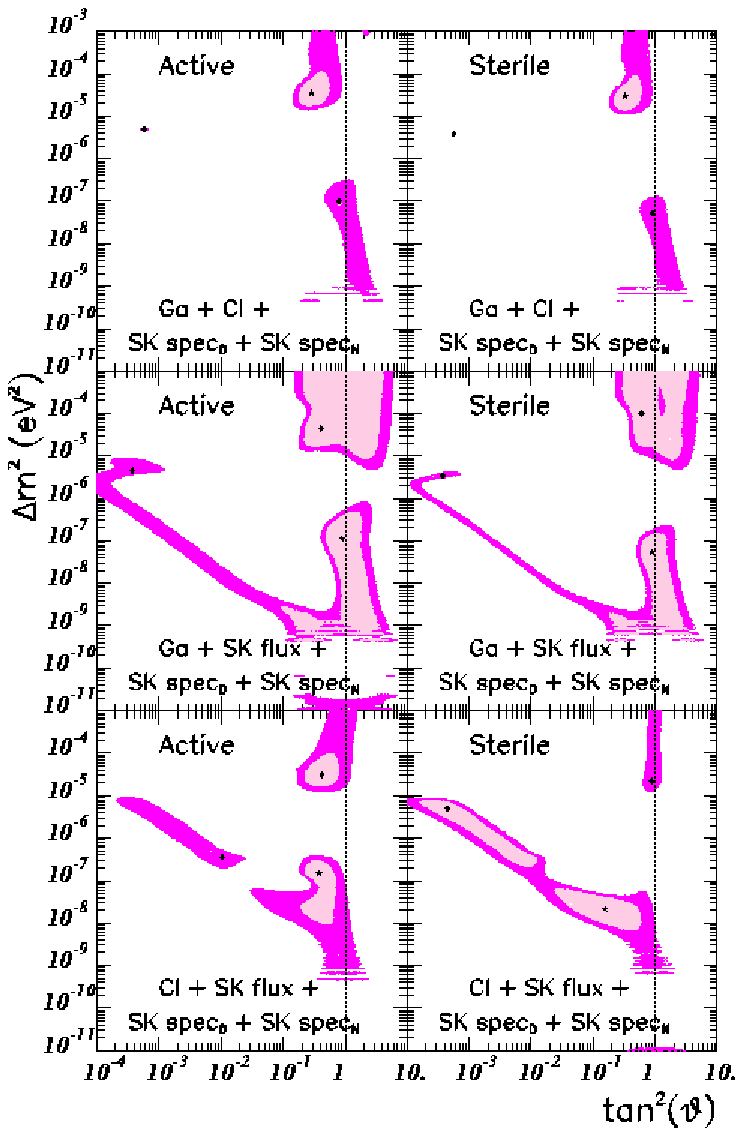}
\end{center}
\vskip -0.9cm
\caption{90 and 99 \% CL allowed regions from the global analysis of  
solar neutrino data where one the fluxes is removed from the data to 
analyze( SK, Cl and Ga respectively). The global minimum is marked with a 
star while the local minima are denoted with a dot.}
\label{two}
\end{figure*}

Finally let's stress that when the allowed parameter space 
for solutions of a physics problem consists of a set of isolated regions, 
the  quality of a given solution cannot be inferred from the size of the 
corresponding region. This fact is nicely illustrated when comparing
the SMA solutions for the active and sterile cases shown in 
Fig.~\ref{global}. Notice that the SMA region for oscillations into sterile
neutrinos is larger than the corresponding one for active neutrinos.
This is because it is defined with respect to the global minimum for 
the sterile oscillation hypothesis (which lies in the SMA region) while the 
SMA region for the active case is defined with respect to the global minimum
for that scenario which lies in the LMA. However when looking at the GOF 
of these two solutions we find that the SMA solution for active case 
provides a better description of the data than the sterile one despite its
region is smaller. In order to put the two scenarios, active and sterile 
in the same footing we must embed them into a common framework as we 
describe in the next section.

\section{Four--Neutrino Oscillations}
In this section we present a brief update of the analysis performed 
in Ref.~\cite{four}. We refer to this publication for further details on the
as well as for the relevant references. Here we simply summarize
the main ingredients.

Together with the results from the solar neutrino  
experiments we have two more evidences pointing out towards the existence of  
neutrino masses and mixing: the atmospheric neutrino data and the LSND 
results. All these experimental results can be accommodated 
in a single neutrino oscillation framework only if there are at least 
three different scales of neutrino mass-squared differences. 
The simplest case of three independent 
mass-squared differences 
requires the existence of a light sterile neutrino, 
{\it i.e.} one whose interaction with 
standard model particles is much weaker 
than the SM weak interaction, 
so it does not affect the invisible Z decay  
width, precisely measured at LEP.

In four-neutrino schemes 
the flavor neutrino fields 
$\nu_{\alpha L}$ 
($\alpha=e,s,\mu,\tau$) 
are related 
to the fields $\nu_{kL}$ of neutrinos with masses $m_k$ 
by the relation 
\begin{equation} 
\nu_{\alpha L} = \sum_{k=1}^4 U_{\alpha k} \, \nu_{kL} 
\qquad 
(\alpha=e,s,\mu,\tau) 
\,, 
\label{mixing} 
\end{equation} 
where $U$ is a $4{\times}4$ unitary mixing matrix, which contains, in general
6 mixing angles (we neglect here the CP phases). 

Existing bounds from negative searches for neutrino oscillations performed
at collider as well as reactor experiments impose severe constrains
on the possible mass hierarchies as well as mixing structures for the
four--neutrino scenario. In particular they imply:

(a) Only two four-neutrino schemes 
can accommodate the results of all neutrino oscillation experiments. 
\begin{figure}[htbp]
\vskip -0.7cm
\begin{center}
\begin{turn}{-90}
\includegraphics[scale=0.3]{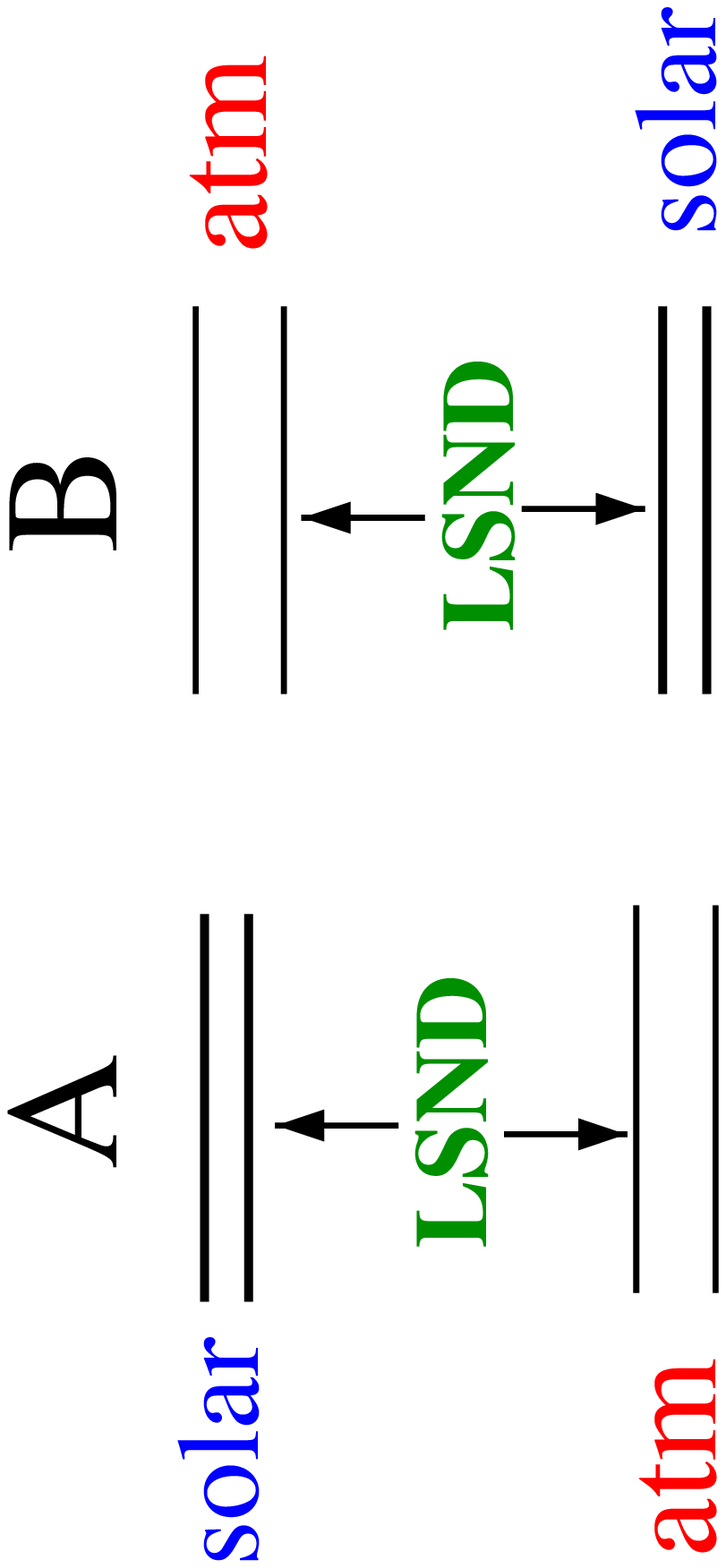}
\end{turn}
\end{center}
\vskip -0.7cm
\end{figure}
In both these mass spectra 
there are two pairs 
of close masses separated by a gap of about 1 eV 
which gives the mass-squared difference 
$ \Delta{m}^2_{{SBL}} = \Delta{m}^2_{41} $ 
responsible for the 
short-baseline (SBL) oscillations observed in the LSND experiment 
(we use the common notation 
$\Delta{m}^2_{kj} \equiv m_k^2 - m_j^2$). 
We have ordered the masses in such a way that 
in both schemes 
$ \Delta{m}^2_{{sun}} = \Delta{m}^2_{21} $ 
produces solar neutrino oscillations 
and 
$ \Delta{m}^2_{{atm}} = \Delta{m}^2_{43} $ 
is responsible for atmospheric neutrino oscillations. 

(b) In the study of solar neutrino oscillations 
only four mixing angles are relevant and the $U$ 
matrix can be written as
\begin{equation} 
U = U_{34} \, U_{24} \, U_{23} \, U_{12} 
\,. 
\label{U-sun} 
\end{equation} 
where 
$\vartheta_{12}$, 
$\vartheta_{23}$, 
$\vartheta_{24}$, 
$\vartheta_{34}$ 
are four mixing angles and we will define 
$ c_{ij} \equiv \cos\vartheta_{ij} $ 
and 
$ s_{ij} \equiv \sin\vartheta_{ij} $. 
 
Since solar neutrino oscillations 
are generated by the mass-square difference 
between $\nu_2$ and $\nu_1$, 
it is clear from Eq.~(\ref{U-sun}) 
that the survival of solar $\nu_e$'s 
mainly depends on the mixing angle 
$\vartheta_{12}$, 
whereas 
the mixing angles 
$\vartheta_{23}$ and $\vartheta_{24}$ 
determine the relative amount of transitions into sterile $\nu_s$ 
or 
active $\nu_\mu$ and $\nu_\tau$. 
Let us remind the reader that 
$\nu_\mu$ and $\nu_\tau$ 
cannot be distinguished in solar neutrino experiments, 
because their matter potential 
and their interaction in the detectors are equal, 
due only to NC weak interactions. 
The active/sterile ratio 
and solar neutrino oscillations in general 
do not depend on 
the mixing angle 
$\vartheta_{34}$, 
that contribute only to the different mixings of 
$\nu_\mu$ and $\nu_\tau$, 
and depends on 
the mixing angles 
$\vartheta_{23}$ 
$\vartheta_{24}$ 
only through the combination 
$\cos{\vartheta_{23}} \cos{\vartheta_{24}}$. 

Therefore, the oscillations of solar neutrinos depend only on 
$\vartheta_{12}$ and the product 
$\cos{\vartheta_{23}} \cos{\vartheta_{24}}$. 
If $\cos{\vartheta_{23}} \cos{\vartheta_{24}} \neq 1$, 
solar $\nu_e$'s can transform 
in the linear combination $\nu_a$ of active $\nu_\mu$ and $\nu_\tau$. 
We distinguish the following limiting cases: 

$\bullet$ $\cos{\vartheta_{23}} \cos{\vartheta_{24}} = 0$ corresponding to the limit of 
pure two-generation 
$\nu_e\to\nu_a$ transitions.
 
$\bullet$ $\cos{\vartheta_{23}} \cos{\vartheta_{24}} = 1$ 
for which we have the limit of 
pure two-generation 
$\nu_e\to\nu_s$ transitions.
 
In the general case of simultaneous $\nu_e\to\nu_s$ and 
$\nu_e\to\nu_a$ 
oscillations 
the corresponding probabilities are given by 
\begin{eqnarray} 
&& 
P_{\nu_e\to\nu_s} 
= 
c^2_{23} c^2_{24} 
\left( 1 - P_{\nu_e\to\nu_e} \right) 
\,, 
\label{Pes} 
\label{Pea} 
\\ 
&& 
P_{\nu_e\to\nu_a} 
= 
\left( 1 - c^2_{23} c^2_{24} \right) 
\left( 1 - P_{\nu_e\to\nu_e} \right) 
\,. 
\end{eqnarray} 
where  $P_{\nu_e\to\nu_e}$ takes the standard two--neutrino oscillation
form (\ref{Pee}) for $\Delta m^2_{21}$ and $\theta_{12}$ but
computed for the modified matter potential  
\begin{equation} 
A 
\equiv 
A_{CC} + c^2_{23}c^2_{24} A_{NC} 
\,. 
\label{A} 
\end{equation} 
Thus the analysis of the solar neutrino data in the
four--neutrino mixing schemes it is equivalent to the two--neutrino
analysis but taking into account that the parameter space is now 
three--dimensional $(\Delta m^2_{21},\tan^2\vartheta_{12}, 
\cos^2{\vartheta_{23}} \, \cos^2{\vartheta_{24}})$. 

We first present the results of the allowed regions in the three--parameter 
space for the global combination of observables. 
In building these regions, for a  
given set of observables,  we compute for any point in the parameter space  
of four--neutrino oscillations 
the expected values of the observables and with those and the corresponding 
uncertainties we construct the function  
$\chi^2(\Delta m_{21}^2,\vartheta_{12},c_{23}^2c_{24}^2)$.  
We find its minimum in the full three-dimensional space. The allowed regions  
for a given CL are then defined as the set of points satisfying  
the condition\\ 
$\chi^2(\Delta m_{21}^2,\vartheta_{12},c_{23}^2c_{24}^2)
-\chi^2_{min}\leq \Delta\chi^2 \mbox{(CL, 3~dof)} $\\
where, for instance, $\Delta\chi^2($CL, 3~dof)=6.25, 7.83, and 11.36 for 
CL=90, 95, and 99 \% respectively. 
In  Figs.~\ref{four} we plot the  
sections of such volume in the plane 
($\Delta{m}^2_{21},\tan^2(\vartheta_{12})$) for different values of 
$c_{23}^2c_{24}^2$.  The global minimum used in the construction of the 
regions 
lies in the LMA region and for pure active oscillations 
value of $c_{23}^2c_{24}^2=0$. 
\begin{figure*}[htbp]
\vskip -0.7cm
\begin{center}
\includegraphics[scale=1.5]{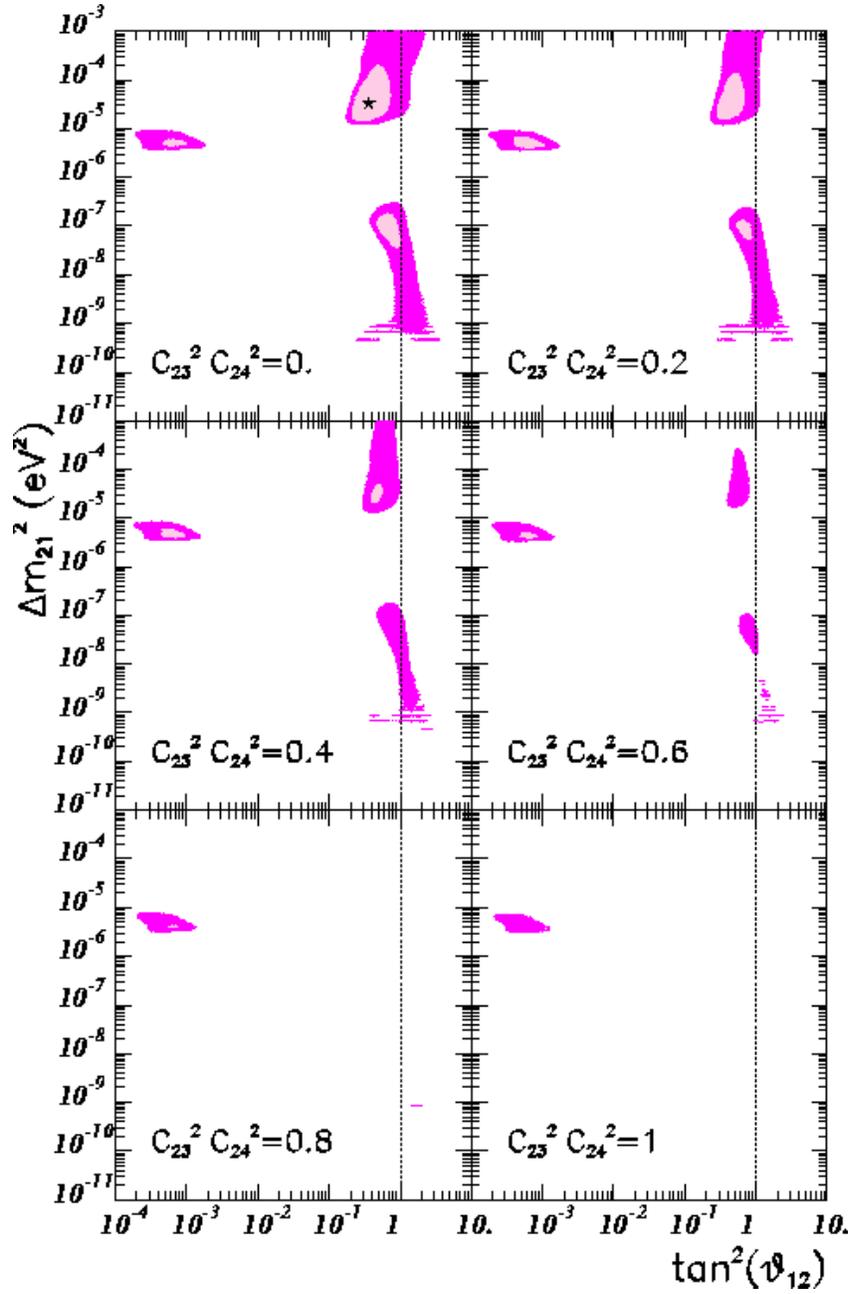}
\end{center}
\vskip -0.7cm
\caption{Results of the global analysis for the allowed regions in   
$\Delta{m}^2_{21}$ and $\tan^2 \vartheta_{12}$  
for the four--neutrino oscillations. 
The different panels 
represent the allowed regions at 99\% (darker) and 90\% CL (lighter)  
obtained as sections for fixed values of the mixing angles  
$c^2_{23} c^2_{24}$ of the three--dimensional volume.
The best--fit point in the three parameter space is  
plotted as a star.} 
\label{four}
\end{figure*}

As seen in  Fig.~\ref{four} 
the SMA region is always a valid solution  
for any value of $c_{23}^2c_{24}^2$. This is expected as  
in the two--neutrino oscillation picture this solution holds both  
for pure active--active and pure active--sterile oscillations.
Notice, however, that the statistical analysis is different: 
in the two--neutrino picture the pure active--active and active--sterile 
cases are analyzed separately, 
whereas in the four--neutrino picture they are taken into account 
simultaneously in a consistent scheme. We see that in this ``unified'' 
framework, since the 
GOF of the SMA solution for pure sterile oscillations is worse than
for SMA pure active oscillations (as discussed in the previous section), 
the corresponding allowed region is smaller as they are now defined 
with respect to a common minimum.
\begin{figure} 
\vskip -0.7cm
\begin{center}
\includegraphics[scale=0.4]{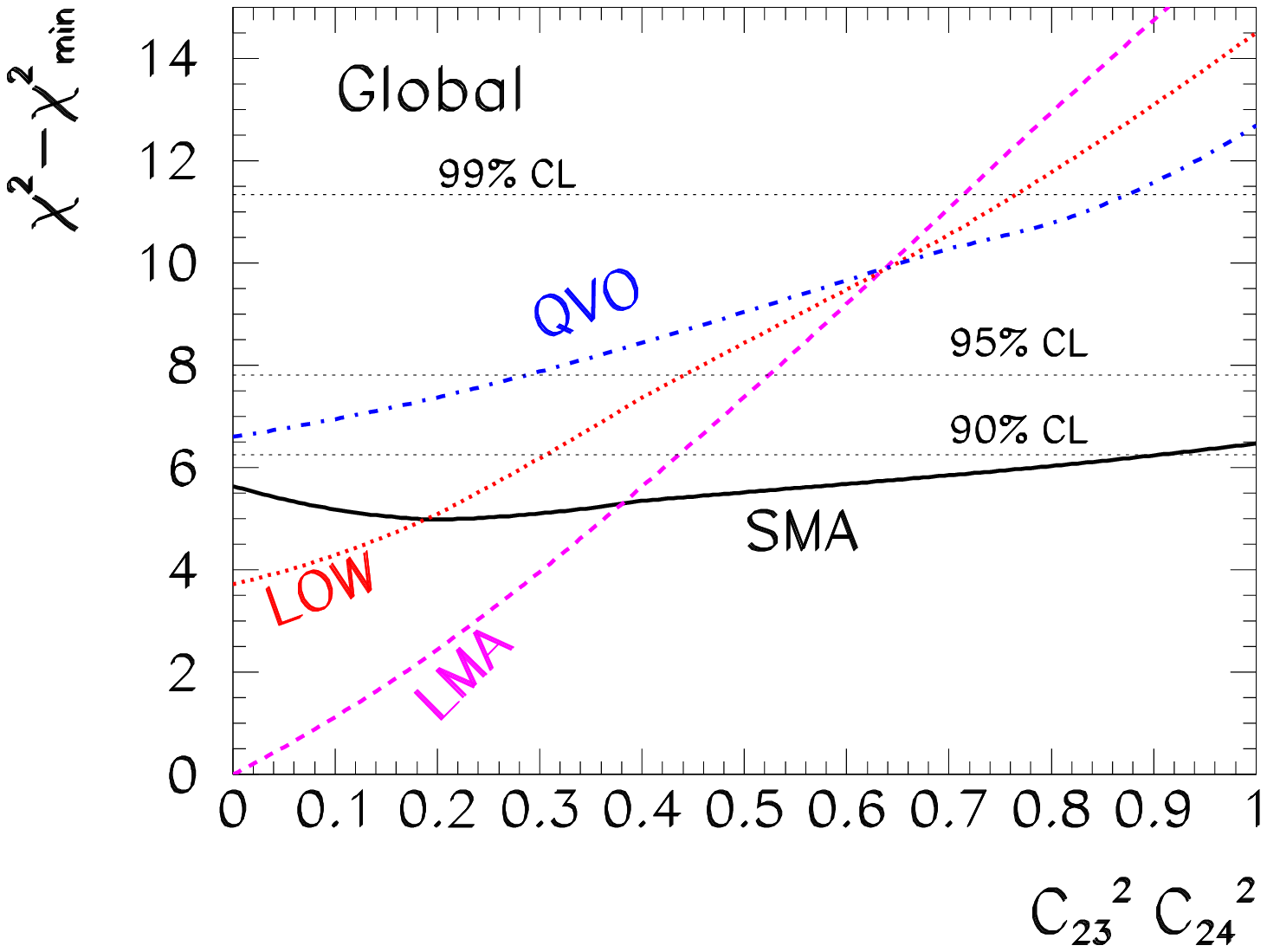}
\end{center}
\vskip -0.7cm
\caption{$\Delta \chi^2$ as a function of the mixing parameter 
$c_{23}^2 c_{24}^2$ for the different solutions SMA (full line), 
LMA (dashed), LOW(dotted) and QVO (dot--dashed) from 
the global analysis of the rates and the day-night spectrum data. 
The dotted horizontal lines correspond to the 90\%, 95 \%, 99\% CL.} 
\label{chi2}
\end{figure}

On the other hand, the LMA, LOW and QVO solutions disappear for 
increasing values of the mixing $c_{23}^2c_{24}^2$. In Fig.~\ref{chi2}  
we show the the values of $\Delta \chi^2$ (with respect to the
global minimum) for the different 
solutions as a function of $c_{23}^2c_{24}^2$. From this figure we
can read the limiting values of $c_{23}^2c_{24}^2$ 
for which
a given solution is allowed in a given CL. We list some of this limits
in Table \ref{limits4}. 
We find that at 95\% CL the LMA is allowed 
for maximal active-sterile mixing $c_{23}^2c_{24}^2=0.5$ while 
at 99\% CL all solutions are possible for this maximal mixing case.
\begin{table}
\caption{Maximum allowed value of $\cos^2_{23} \cos^2_{24}$ at 90\%, 95\%, 
and 99 \% CL for the different solutions to the solar neutrino problem.}  
\label{limits4}
\begin{center}
\begin{tabular}{c|cccc}
CL  & SMA & LMA & LOW & QVO \\ \hline
90  & 0.9 & 0.44 & 0.3 & forbidden \\
95  & all & 0.53 & 0.44 & 0.28 \\
99  & all & 0.72 & 0.77 & 0.88 
\end{tabular}
\end{center}
\end{table}

Other authors have recently performed fits of the atmospheric neutrino data
in the framework of the two four-neutrino schemes A and B
\cite{Yasuda-Lisi}.
We expect that in the future a combined fit of solar and atmospheric
neutrino data will allow to constraint 
further the mixing of four neutrinos.

\section{Summary and Conclusions}

We have studied the status of the solutions of the solar 
neutrino problem in terms of oscillations of $\nu_e$ 
into active or sterile neutrinos in the framework of two--neutrino and
four--neutrino mixing, after a global analysis of  
the full data set corresponding to the latest SK 
data on the total event 
rate, their day--night dependence and the recoil electron
energy spectrum, together with the data from Chlorine and Gallium 
experiments presented at the $\nu$-2000 conference. 
In particular we have shown the possible  solutions in the full parameter 
space for oscillations including both MSW and vacuum, as well as 
QVO and matter effects for mixing angles in the second octant 

Our main conclusions from the two--neutrino oscillation analysis 
are the following:
\begin{itemize}
\item[(a)] From the statistical point of view,  
all solutions for oscillations into active neutrinos:  LMA, LOW, SMA, QVO 
solutions are acceptable since they all provide a reasonable GOF 
to the full data set. 
\item[(b)] The same holds for the SMA solution, which is the best
possible solution for oscillations into sterile neutrinos. 
\item[(c)] LMA and LOW-QVO solutions for oscillations into 
active neutrinos seem slightly favoured over SMA solutions for oscillations 
into active or sterile neutrinos due to the flatter spectrum of SK.
\item[(d)] Oscillations into sterile neutrino provide a slightly worse
fit. 
\item[(e)] When allowing for a free $^8$B normalization weakens the conclusion 
(c).
\item[(f)] Removing the total rate measured at Chlorine or SK from the
analysis weakens conclusion (d) and straightens conclusion (c).
\end{itemize}

For the analysis of four-neutrino oscillations, which allows for 
oscillations into a state which is a 
combination of active and sterile neutrino, we find that 
the SMA region is always a valid solution  while 
the LMA, LOW and QVO solutions disappear for 
increasing values of the additional mixing $c_{23}^2c_{24}^2$. 
However we find that at 95\% CL the LMA is still allowed 
for maximal active-sterile mixing $c_{23}^2c_{24}^2=0.5$ while 
at 99\% CL all solutions are possible for this maximal mixing case.


\begin{thebibliography}{9}

\bibitem{bp98} J. N.~Bahcall,  S.~Basu and M. H.~Pinsonneault,
Phys.~Lett. {\bf B433}, 1 (1998). 

\bibitem{chlorine} B. T. Cleveland {\it et al.},
 Astrophys. J. {\bf 496}, 505 (1998); 
 R. Davis, Prog. Part. Nucl. Phys. {\bf 32}, 13 (1994).
 
\bibitem{sage} SAGE Collaboration, J. N. Abdurashitov {\it et al.},
 Phys. Rev. {\bf C60}, 055801 (1999); Talk by V. Gavrin at
 Neutrino 2000, Sudbury, Canada, June 2000
  (http://{\-}nu2000.{\-}sno.{\-}laurentian.{\-}ca). 
 
\bibitem{gallex} GALLEX Collaboration, W.~Hampel {\it et al.},
 Phys. Lett. {\bf B447}, 127 (1999).

\bibitem{gno} Talk by E. Belloti at 
 Neutrino 2000, Sudbury, Canada, June 2000
  (http://{\-}nu2000.{\-}sno.{\-}laurentian.{\-}ca). 

\bibitem{superk} Super--Kamiokande Collaboration, Y. Fukuda {\it et al.},
 Phys. Rev. Lett. {\bf 81}, 1158 (1998); Erratum {\bf 81}, 4279 (1998); 
 {\bf 82}, 1810 (1999); {\bf 82}, 2430 (1999);
  Y. Suzuki, Nucl. Phys. B (Proc. Suppl.) {\bf 77}, 35 (1999).

\bibitem{suzuki}
 Talk by Y. Suzuki  at 
 Neutrino 2000, Sudbury, Canada, June 2000
  (http://{\-}nu2000.{\-}sno.{\-}laurentian.{\-}ca). 
  
\bibitem{sno} A. B.~McDonald,
 Nucl. Phys. B (Proc. Suppl.) {\bf 77}, 43 (1999);
 SNO Collaboration, Physics in Canada {\bf 48}, 112 (1992);
 SNO Collaboration, nucl-ex/9910016.


\bibitem{Glashow:1987jj}
V.N.~Gribov and B.M.~Pontecorvo, Phys. Lett. {\bf 28B}, 493 (1969);
V. Barger, K. Whisnant, R.J.N. Phillips, Phys. Rev. {\bf D24}, 538
(1981); S.L.~Glashow and L.M.~Krauss, Phys. Lett. {\bf 190B}, 199
(1987).

\bibitem{msw}
S.P. Mikheyev and A.Yu. Smirnov, Sov. Jour. Nucl. Phys. 
42, 913 (1985); L.\ Wolfenstein, Phys.\ Rev.\ {\bf D17}, 2369 (1978).

\bibitem{LZ} 
S.~T.~Petcov, Phys. Lett. {\bf B200}, 373 (1988); 
P.~I.~Krastev and S.~T.~Petcov, Phys. Lett. {\bf B207}, 64 (1988).


\bibitem{BP00} 
http://www.{\-}sns.{\-}ias.{\-}edu/\~\,{\-}jnb/{\-}SNdata/{\-}Export/{\-}BP2000; J. N. Bahcall, S. Basu 
and M. H. Pinsonneault, Astrophys. J. {\bf 529}, 1084 (2000).

\bibitem{PREM} 
A.~M.~Dziewonski and D.~L.~Anderson, Phys. Earth Planet. Inter. 
{\bf 25}, 297 (1981).

\bibitem{panta} J. Pantaleone, Phys. Lett. {\bf B251},618 (1990); S. Pakvasa
 and J. Pantaleone, Phys. Rev. Lett. {\bf 65}, 2479 (1990)

\bibitem{fried} A. Friedland,  Phys. Rev. Lett. {\bf 85}, 936 (2000).
 
\bibitem{threev} G. L. Fogli, E. Lisi, D. Montanino and A. Palazzo,
 hep-ph/0005261; A.M. Gago, H. Nunokawa, R. Zukanovich Funchal, 
 hep-ph/0007270. 
\bibitem{maxmix} M.C. Gonzalez-Garcia, C. Pe\~na-Garay, 
Y. Nir, A. Yu. Smirnov, hep-ph/00007227.

\bibitem{dGFM} A. de Gouvea, A. Friedland and H. Murayama, hep-ph/0002064.

\bibitem{GGPG}
 M. C. Gonzalez-Garcia, C. Pe\~na-Garay, Phys. Rev. {\bf D62}, 031301 (2000).

\bibitem{three} G. L. Fogli, E. Lisi, A. Maronne and G. Scioscia,
 Phys. Rev. {\bf D59}, 033001 (1999).

\bibitem{gonzalez} M. C. Gonzalez-Garcia, P. C. de Holanda,
C. Pe\~na-Garay, J. W. F. Valle,  Nucl. Phys. {\bf B573}, 3 (2000).


\bibitem{four} C. Giunti, M. C. Gonzalez-Garcia, C. Pe\~na-Garay, 
Phys. Rev. {\bf D62}, 013005 (2000).

\bibitem{Yasuda-Lisi}
O. Yasuda, hep-ph/0006319; 
E. Lisi, Talk presented at Neutrino 2000, Sudbury, Canada, June 2000
  (http://{\-}nu2000.{\-}sno.{\-}laurentian.{\-}ca).
\end{thebibliography}
\end {document}